\documentstyle{article}
\title{BD+30$^\circ$3639: The infrared spectrum during
post-AGB stellar evolution}

\author{Ralf Siebenmorgen $^{1,2}$, Albert A. Zijlstra$^2$ and
Endrik Kr\"ugel$^3$\\
\\
\\
$^1$ ESTEC\\
Science Astrophysics Division\\
ISO Science Operation Team\\
Postbus 299\\
2200 AG Noordwijk\\
The Netherlands\\
\\
$^2$ European Southern Observatory\\
Karl-Schwarzschildstr. 2\\
D-85748 Garching bei M\"unchen\\
Federal Republic of Germany\\
\\
$^3$ Max-Planck-Institut f\"ur Radioastronomie \\
Auf dem H\"ugel 69 \\
W-5300 Bonn \\
Federal Republic of Germany\\
\\
\\
accepted for publication in MNRAS}
\newpage
\begin{document}
\maketitle
\newpage


\def\wisk#1{\ifmmode{#1}\else{$#1$}\fi}
%
%
\def\k{\,{\rm K}}
\def\khz{\,{\rm kHz}}
\def\mhz{\,{\rm MHz}}
\def\ghz{\,{\rm GHz}}
\def\kmsmpc{{\rm\,km\,s^{-1}\,Mpc^{-1}}}                       
\def\jy{\,{\rm Jy}}                                            
\def\kpc{\,{\rm kpc}}                                          
\def\pc{\,{\rm pc}}
\def\yr{\,{\rm yr}}                                            
\def\mjy{\,{\rm mJy}}
\def\magg{\,{\rm mag}}                                         
\def\peryr{\,{\rm yr^{-1}}}
\def\cm{\,{\rm cm}}
\def\cmd{\,{\rm cm^{-3}}}
\def\angstrom{\,{\rm\AA}}
\def\msolar{\,{\rm M_\odot}}
\def\lsolar{\,{\rm L_\odot}}
\def\mum{\mu{\rm m}}
%
%
\def\hbeta{\wisk{ H\beta}}
\def\lhbeta{\wisk{ L_\hbeta}}
\def\fhbeta{\wisk{ F_\hbeta}}
\def\lyalpha{\wisk{ Ly\alpha}}
\def\halpha{\wisk{ H\alpha}}
\def\ebv{\wisk{ E_{B-V}}}
\def\teff{\wisk{ T_{eff}}}
\def\lstar{\wisk{ L_\star}}
\def\tstar{\wisk{ T_\star}}
\def\lsol{\wisk{\,\rm L_\odot}}
\def\msol{\wisk{\,\rm M_\odot}}
\def\rsol{\wisk{\,\rm R_\odot}}
\def\tsol{\wisk{\,\rm T_\odot}}
\def\kms{\wisk{\,\rm km\,s^{-1}\,}}                    
\def\wcm{\wisk{\,\rm W\ cm^{-2}\,}}                    
\def\wcmmu{\wisk{\,\rm W\,cm^{-2}\,\mu m^{-1}\,}}      
\def\cmcub{\wisk{\,\rm cm^{-3}\,}}                     
\def\tenpow#1{\wisk{\,\rm 10^{#1}\,}}                  
\def\timtenpow#1{\wisk{\,\rm\times10^{#1}\,}}          
\def\jyb{\wisk{\,\rm Jy\ beam^{-1}\,}}                 
\def\mjyb{\wisk{\,\rm mJy\ beam^{-1}\,}}               
%
%
\def\gt   {$\!$\hbox{\tt >}$\!$}
\def\lt   {$\!$\hbox{\tt <}$\!$}
\def\oversim#1#2{\lower0.5pt\vbox{\baselineskip0pt \lineskip-0.5pt
     \ialign{$\mathsurround0pt #1\hfil##\hfil$\crcr#2\crcr\sim\crcr}}}
\def\gsim{\mathrel{\mathpalette\oversim>}}    
\def\lsim{\mathrel{\mathpalette\oversim<}}    
\def\deg{\wisk{^{\rm o}}}                                
\def\degpt#1{#1\wisk{^{\rm o}}}                         
\def\arcmin#1{#1\wisk{^{\prime}}}                        
\def\arcsec#1{#1\wisk{^{\prime\prime}}}                  
\def\decdeg#1.#2{\wisk{#1^{\,\rm o}\bck.\,#2}}
\def\sdecdeg#1.#2{\wisk{#1^{\phantom{\,\rm o}}\bck.\,#2}}
%

%
%

\begin {center}
{\Large \bf BD+30$^{\circ}$3639: The infrared spectrum during
post-AGB stellar evolution}
\end {center}
\vskip2.0truecm
\begin{abstract}

We present a radiative-transfer calculation which reproduces the
infrared spectrum of the planetary nebula BD~+30$^{\circ}$3639. We
calculate the transfer process through absorption and scattering in a
spherical-symmetric multi-grain dust shell. The emission of
transiently heated particles is taken into account, as well as
polycyclic aromatic hydrocarbons.  We obtain an acceptable fit to most
of the spectrum, including the PAH infrared bands.  At submillimetre
wavelengths the observed emission is larger than the model predicts,
indicating that large dust conglomerates (``f{}luffy grains'') may be
needed as an additional constituent. The fit favour a distance of
$\ge 2 \,$kpc, which implies that BD~+30$^\circ$3639 has evolved from
a massive progenitor of several solar masses.  A low dust-to-gas mass
ratio is found in the ionised region.  The calculations yield an
original mass-loss rate of $2\times10^{-4}
\msolar \peryr$ on the Asymptotic Giant Branch. Using this mass-loss
rate, we calculate how the infrared spectrum has evolved during the
post-AGB evolution.  We show in particular the evolution of the IRAS
colours during the post-AGB evolution.
\\
\end{abstract}

\baselineskip 11pt

\section{Introduction}

Low-mass stars in their latest stages of evolution play an important
role for their host galaxies (e.g. bolometric luminosity, chemical
evolution). By now it seems to be established that all stars with a
main-sequence mass of 0.8--8$\msolar$ form white dwarfs, evolving via
the giant and asymptotic giant branch (AGB) through the planetary
nebulae (PN) phase (Iben \& Renzini 1983).  On the AGB the stars
undergo a large loss of mass and a large amount of dust is created,
enriching the interstellar medium.  While there exists a substantial
amount of theoretical models describing the thermo-nuclear evolution
of the cores of such stars (Weidemann \& Sch\"onberner 1989), the
physical mechanism of mass loss from these stars remains unclear.
Present understanding is to a great extent based on theoretical
conjectures whereas observational data are scarce.

After the mass loss ends, the star rapidly becomes hotter, and quickly
begins to ionize the expanding envelope. At present, the study of
objects in the earliest phase of ionization (being in transition to
planetary nebula) gives the best tool to study the preceding mass
loss.  In this paper, we therefore present a detailed study of a young
planetary nebula, BD~+30$^\circ$3639.  The infrared spectrum of this
source has recently been modeled by Hoare et al. (1992, hereafter
HRC). We use a dust model with additional components as compared to
these authors; on the other hand, HRC also model the
photo-ionisation which we do not include.  After the dust parameters
are derived, we develop a simple evolutionary scenario to describe how
its infrared spectrum may have evolved.  We have to solve two basic
problems: i) The objects are highly embedded in an optically thick
dusty envelope and the equation of radiative transfer has to be
properly handled. ii) A more or less realistic description of the
nature of circumstellar dust particles is required as well (Barlow
1993).

This article is structured as follows. In Section 2 we briefly
describe a radiative-transfer model, which takes transiently heated
particles and large molecules as the proposed carrier of the broad
near- and mid-infrared emission features into account.  We than fit
the spectral energy distribution of the well-observed carbon-rich
planetary nebula (PN) BD~+30$^\circ$3639 (Section 3). The fit
parameters are discussed in Section 4. The calculated spectra at
several stages of the preceding post-AGB evolution are presented in
Section 5. We conclude with a summary of our findings in Section 6.

\section{The dust--radiative-transfer model}

\subsection{The dust model}

The dust model used in the present calculations is that of
Siebenmorgen \& Kr\"ugel (1992, henceforth SK). This model succesfully
reproduces observations ranging from near-infrared to millimeter
wavelengths, including narrow-band data. The grain properties have
been established by fitting extinction curves towards different lines
of sight (e.g. Fitzpatrick \& Massa 1988, Cardelli et al. 1989, Mathis
1990). The dust model is governed by a grain size distribution
($n(a)\propto a^{-q}$) with radii between approximately 2500\AA \/
down to molecular sizes of about 5\AA \/ ($\sim$ 25 atoms) in which we
distinguish three different populations of dust particles:
\\
i) Large particles with sizes $a \geq 100$\AA, $q = 3.5$, causing the
far-infrared/submillimeter emission.  We call these particles large
because they have sufficient heat capacity that we can neglect
the quantum-statistical behavior of photon--grain interactions.
Since in this paper we only discuss carbon-rich environments which do
not exhibit the Si-O stretching and bending modes at $9.7\mum$ and $18
\mum$, silicate grains are not included.  We have used two
 different sets of optical constants: (1) constants given by Draine \&
Lee (1984, 1987) for grains with graphite structure; (2)
constants for amorphous carbon for $300 \AA \leq \lambda \leq 0.105
\mu m$ published by Rouleau \& Martin (1991) and for $0.105 <\lambda
\leq 800 \mu m$ by Preibisch et al. (1993). Compilation (2)
is based on optical constants from Bussoletti et al. (1987) and Blanco
et al.  (1991) and re--analyzed by taking shape and clustering effects
into account.
\\
ii) Small graphite grains with sizes 10\AA$ \leq a < 100$\AA, $q = 4$,
emitting predominately in the mid-infrared.  These particles show
temperature fluctuations after individual energy-absorption events
(e.g. with photons, electrons); to calculate their emission spectrum
one has to consider multi-photon events.  We use properties such as
abundance, bulk density, absorption efficiencies and enthalpy of these
small graphites as given in SK.  Small graphites are at present the
most popular candidates to explain the 2200\AA\ extinction bump
(Draine 1988, 1989; Sorrel 1989).
\\
iii) Small PAHs ($\sim 25$ atoms) and larger PAH clusters ($\sim 250$
atoms). L\'eger \& Puget (1984) and Allamandola et al.  (1985) have
argued that these large isolated molecules explain the near- and
mid-infrared emission bands. A review of the whole family of emission
features between 3--14$ \mu$m can be found in, e.g.,  Allamandola et
al. (1989a), Puget \& L\'eger (1989) and Tokunaga et al.  (1991). From
the ratios of different emission bands a temperature of the emitting
species can be derived; if one adopts a model for the heat capacity, a
size can be estimated as well.  By doing this for the $3.3 / 3.4$ and
$3.3 / 11.3$ band ratios, de Muizon et al. (1990) found indications
that PAHs also show a size distribution.  The band ratios of the C--H
to the C=C vibrational modes will be affected by the number of H atoms
attached to the peripheral rings in a PAH.  Therefore the
hydrogen-to-carbon atom ratio is introduced, called the
hydrogenation parameter. Emission coefficients of model PAHs are
given by L\'eger et al. (1989a), d'Hendecourt et al.  (1989) and more
recently by de Muizon et al.  (1990). PAHs absorb predominately in the
far-UV and, as demonstrated in laboratory studies (L\'eger et al.
1989b, Joblin et al. 1992), an ensemble of these molecules can be
responsible for the non-linear rise towards the UV in the extinction
curve. Absorption efficiencies are calculated as described in SK for
$\lambda \geq 912 $\AA. For shorter wavelengths we simply scaled the
absorptivity to a small graphite particle of similar
size (calculated with Mie Theory).
In addition we have compared  our method (for $\lambda \geq 912$\AA)
with the analytical fit given by Desert et al.  (1990) and
found a very good agreement.\\

\subsection {The radiative transfer model}

The solution of the radiative transfer problem for spherically
symmetric dusty objects containing quantum-heated particles and
including scattering by grains is described by Siebenmorgen et
al. (1992).  The method outlined for the photo-destruction of
transiently heated particles is given by Siebenmorgen (1993). The
procedure does not take into account the heating of the dust by the
emission of ionized gas, nor does it treat any of the cooling lines
observed between UV and far-IR wavelengths.

Important is the neglect of Ly$\alpha$ radiation: The HII gas emits
Ly$\alpha$ photons and Lyman-continuum photons from recombination to
the ground state, both of which are trapped in the HII region.  Under
certain conditions, Ly$\alpha$ radiation can dominate the heating of
the grains and even cause their destruction. This happens especially
if the HII region is heavily dust depleted, because the energy input
from Ly$\alpha$ per grain is proportional to the number of Lyman
continuum photons of the star that lead to ionization divided by the
mass of the dust in the HII region. Since this is not presently
included in the model, it is important to estimate the effects of this
radiation, to understand the limitations of the present calculations.

In order to estimate the effect of especially the Ly$\alpha$
radiation, we attempted to calculate the structure of the ionized gas
using the on-the-spot approximation. However, the location of the
ionization front could not be fixed sufficiently accurately: this
resulted in violation of the overall energy equation.  Ignoring this
problem, the results of this calculation did not indicate significant
differences with those obtained without ionization and we did not
find a qualitative change of the contribution of small graphites,
PAHs or large-grain component to the overall spectral energy
distribution. However, we cannot rule out that the Ly$\alpha$
radiation leads to significant PAH destruction. This can be important
for a detailed discussion of the spatial distribution of the PAH band
emission (see Section 4.3).

\section{The infrared spectrum of BD~+30$^\circ$3639}

\subsection{Basic parameters}

The program that has been developed to handle the theoretical
framework outlined in the previous section allows flexibility in the
specification of the various grain populations, the heating source,
the dust density distribution, and other physical parameters.  We
adopt the dust model as discussed above. The heating source is
described as a blackbody with an adopted stellar temperature $T_{*}$
and bolometric luminosity. The density distribution can be arbitrary
but is defined here by means of power laws $n(r) \propto r^{\beta}$.

\subsection{ Available data}

BD~+30$^\circ$3639 is one of the best observed planetary nebulae in
the infrared.  All the PAH features are well observed
(e.g. Allamandola et al. 1989b), including the $6.2$ and $7.7 \mu$m
bands which are not observable from the ground. New observations were
communicated to us prior to publication (Schutte and Tielens
1993). Mid-infrared photometry obtained at the Kuiper Airborne
Observatory (Moseley and Silverberg 1985), show that there is no
$30\mu$m emission bump, as seen in the other carbon-rich planetary
nebulae NGC~7027 and IC~418. This feature is usually attributed to MgS
grains or mantles (Goebel \& Moseley 1985).  Since our model does not
include this feature, BD+30$^\circ$3639 gives the best test case.
Submillimetre observations have recently been published by HRC. These
data points are crucial to constrain the density distribution at large
distance from the central star.  At wavelengths $ \geq 800\mu$m the
emission is mostly due to the radio free--free emission: the $6\,$cm
radio flux of $620\,$mJy (Basart and Daub 1987, Masson 1989)
extrapolated to $800\mu$m predicts a flux density of $400\,$mJy, which
accounts for most of the observed value of $550\pm34\,$mJy. At shorter
wavelengths the emission is mostly due to dust.  Finally, the nebula
has been imaged at several mid-infrared bands (Hora et al. 1993,
1990). The model calculates the radial flux distribution, which can be
compared with these observations.

\subsection{ Distance and luminosity}

Two important input parameters of the model are the distance to the
nebula and the luminosity of the central star. In our model, the total
luminosity can be determined as one of the fit parameters, after a
value for the distance has been chosen.

The distance to BD~+30$^{\circ}$3639 is controversial.  HI absorption
features indicate a distance of $\gsim 2.5 \,$kpc, but the low
extinction
($E_{\rm B-V}$=0.24, both from the ratio of the \hbeta\ flux over the
radio flux density and from the depth of the 2200\AA\ feature
towards the star, Pottasch 1984) indicates a distance of about $0.6
\,$kpc (Gathier et al.  1986).  Masson (1989) has measured the
expansion
of the nebula from radio images, from which he derives a large
distance of $2.8^{+48}_{-1.2} \,$kpc.  Hajian et al. (1993), using the
same method, find a distance of $2.68\pm0.81\,$kpc, in agreement with
Masson.  We have tried to fit the infrared spectrum using several
different distances.  For distances $\leq 2\,$kpc, the model required
an inner radius of the circumstellar shell significantly smaller than
the observed value (which is $1.5\arcsec$--$2\arcsec$: Taylor et al.
1987, Masson 1989, Hora et al. 1990, 1993).  This could be solved if
we assume that the abundance of the very small grains is much higher
than in the interstellar medium, although in the fit this parameter
appears to be well constrained.  At larger distances, the observed
inner radius can be fitted by fine-tuning of the stellar luminosity
without requiring changing the dust composition.

We adopt a distance of $2\,$kpc, the same distance as taken by HRC.
This is close to the lowest value allowed by the radio expansion
measurements of Hajian et al., and allows us to fit the spectrum
without changing the abundance of the very small grains with respect
to the interstellar medium.

An approximate luminosity can be obtained from the optically thin
radio luminosity.  The number of ionizing photons required to produce
the observed radio flux can easily be calculated (e.g.  Zijlstra and
Pottasch, 1989), and for a given temperature of the central star
converted to a total luminosity e.g. using a black-body atmosphere.
Because direct absorption of ionizing photons by dust is ignored in
this method, only a lower limit to the stellar luminosity is obtained.
Using the known temperature of the central star of BD+30$^\circ$3639
of $30\,000 \,$K (Pottasch 1984), we find a lower limit to the
luminosity of $7\times10^3\lsolar$ at a distance of $2\,$kpc, adopting
a black-body atmosphere.  More realistic atmosphere models would
predict a somewhat higher luminosity.  In the present model, we find
that the best fit required a luminosity of $1.3\times 10^4\lsolar$;
this is consistent with the limits derived above.

In principle the luminosity is determined by the temperature and
optical magnitude of the star. However, the temperature of
planetary-nebulae central stars is not generally known to sufficient
accuracy for this to give useful results, especially because of the
$T^4$ dependency. Also, the central star of BD~+30$^{\circ}$3639 is of
[WC] type for which a black body is probably not a good approximation.
The luminosity derived here is therefore probably the best
available. As shown later, the K-magnitude of the star is reasonably
well reproduced by the model.

\subsection{ The density distribution and dust composition}

We find that a single constant-density component ($\beta = 0$) cannot
reproduce the infrared spectrum.  By using a constant density for the
inner region, and a $r^{-2}$ component in the outer region, a
satisfactory fit is obtained. The use of a $r^{-2}$ component is known
to give a good approximation to the global density profile of
planetary nebulae (e.g. Taylor et al. 1987); such a structure can be
formed through a phase of constant mass loss. At smaller scales
deviations can be expected from such a distribution: many planetary
nebulae are known to show bipolar structures, which are evidence for
asymmetric mass loss, and are affected in the inner parts by shocks
and pressure inequilibrium arising from the ionization of the nebula.
BD~+30$^{\circ}$3639 also shows enhanced density around the equatorial
plane (which lies north--south, e.g.  Basart and Daub 1987).  However,
the density contrast between pole and equator appears to be much less
than seen in many other PN.  Due to the combined effects of dust
destruction and dust removal by radiation pressure, the dust density
distribution may not perfectly follow that of the gas.  The density
distribution is a simplification, but it probably gives a reasonable
approximation to the actual global structure.

The density structure is parameterized as:
\\
\\
$ \rho = \cases {A_1 & for \ $R_{\rm i}<r<R_2$\cr
       A_2 r^{-2} \quad & for \  $R_2<r<R_{\rm o}$\cr} $
\\
\\
\noindent
where $\rho$ is the hydrogen density.  The parameters are:
the inner and
outer radii $R_{\rm i}$ and $R_{\rm o}$, the radius $R_2$ where the
density changes from uniform to wind-like, and the two density
constants $A_1$ and $A_2$.  The value chosen for the inner radius
affects the slope of the near-infrared spectrum, because this
emission is partly due to the hottest large grains
which are found near the
inner boundary. (The PAHs and small grains also affect the
near-infrared continuum.)
The outer radius mainly affects the submillimetre
emission. We will see that the model does not well constrain this
parameter.  The value of $R_2$ as well and the ratio of the densities
in the two components ($A_1 / (A_2\,{R_2}^2)$) affect the shape of the
spectrum around 50$\mu$m. We use the value of $A_1 = 2\times10^4\,$
cm$^{-3}$ (Pottasch 1984).  HRC find $1.56\times10^4\,$
cm$^{-3}$: using this value would slightly change the
dust-to-gas ratio
but not the derived dust parameters.

Finally, the main constituent of the dust has to be chosen. Because
BD~+30$^{\circ}$3639 is a carbon-rich nebula, only carbon grains are
expected to be present: all available oxygen is quickly locked into CO
during the mass-loss event, and dust is only formed from the remaining
carbon atoms.  In recent years attention has shifted from graphites to
amorphous grains for the large-grain component of carbon dust (Draine
1989, Sorrell 1989, SK).  We find that the mid-infrared emission of
BD+30$^{\circ}$3639 cannot be fitted as well with large graphite
grains, although this might be improved by using a more complicated
density distribution.  A somewhat better fit in the submillimetre is
obtained with the amorphous grains of Preibisch et al.  HRC also
prefer amorphous grains because of the submillimetre emission.  Below
we will present fits for both graphite and amorphous grains. For the
time evolution calculated in Section 5 we have only used models with
amorphous carbon as the large-grain component. The relative abundances
of the different components (large grains, small graphites and two
kinds of PAHs) are determined separately in the model.  It should be
noted that we assume that the dust composition does not depend on
position in the nebula.  This may be an important simplification.

The band ratios of the PAH features at 3.3, 6.2, 7.7, 8.6 and
11.3$\mu$m are governed by the size and the hydrogenation of the PAHs.
Because the bands at 3.3, 8.6 and 11.3$\mu$m are due to aromatic C--H
modes, whereas the other two are caused by C=C aromatic stretching
bands, the hydrogenation will determine the ratio between these lines.
The ratio between the 3.3 and 11.3$\mu$m features is often interpreted
as due to the size of the PAHs (Allamandola et al. 1989a, de Muizon et
al. 1990), whereas in the SK model the small PAHs emit strongly at
$3.3\mu$m and the PAH clusters determine the 11.3$\mum$ feature.

Figure 1 shows the effect of the different dust components on a
calculated spectrum. The large grains, due to their low temperature,
dominate in the far infrared. The very small grains contribute to both
the mid- and near-infrared, the latter because of their very high
temperature excursions. The PAHs cause the band features and part of
the continuum emission between 6 and 13$\mu$m.

\subsection{ Determining the fit parameters}

In the fitting procedure the distance was kept constant to $2\,$kpc,
which, as mentioned above, should be considered as a lower limit.  We
then varied the luminosity and the density parameters, while keeping
the dust parameters constant, in order to find a parameter set which
would approximately reproduce especially the long-wavelength part of
the spectrum.  After a reasonable fit was obtained, we varied the
secondary parameters, such as the relative abundance of each of the
dust components, the size distribution of the PAHs and the
hydrogen-to-carbon ratio of the PAHs, in order to fit the
short-wavelength continuum and the near- and mid-infrared features.
Finally, small adjustments were made to all parameters to arrive at
the ``best fit''.

The resulting fit is shown in Figure 2 using amorphous and Figure 3
using graphite grains.  Detailed parameters of the fit are presented
in Table 1 for both models.  In the upper panel of each Figure the
full spectrum is presented. It can be seen that not all observed data
points are consistent with each other. Some of the presented data
points have been read off from published spectra (Moseley and
Silverberg 1985, Allamandola et al.  1989b, Hora et al.  1990, 1993,
the IRAS LRS spectrum: IRAS science team 1986, Russel et al. 1977);
others come from the compilation by Bentley et al. (1984).  The
published errors are generally smaller than the size of the symbols,
although some points may have larger uncertainties. The J and H data
come from the listing of Acker et al.  (1992); the original
measurements are from Pe\~na and Torres-Peimbert (1987).  The K-data
are from Hora et al.  (1993): the lower data point is the star only,
the upper nebula plus star.  Data points read of from the spectrum
published by Russel et al. are shown by the triangles: at K this
spectrum falls below the data from Hora et al.  One data point
corresponding to the [SIII] emission line at $18.68\mu$m is not shown.
The IRAS flux densities at 25, 60 and 100$\mu$m lie above the Kuiper
Airborne Observatory measurements.  This could be due to aperture
effects.  The IRAS scale could also be slightly wrong, since it is
derived from model fits to the zodiacal emission. New observations at
several apertures are desirable to remove this inconsistency.  The
lower panel shows the new data of Schutte and Tielens (1993).

\section{ Discussion}

\subsection{ Overview }

Figure 2 shows a fit which is accurate over most of the spectrum to
within $\leq 20\%$. This is a good result, especially considering that
our model is not a prefect representation of the nebula: the
incomplete gas--grain interaction (e.g. neglected photoionization),
the simple density model and the significant uncertainties in the
optical constants of the dust properties. The fit in Figure 3, which
uses graphite grains, does not well represent the shape of the
continuum around $40\mu$m.  We note that amorphous carbon grains
clearly give the better fit, but  this is insufficient to
decisively choose against large graphite grains.  The general
expectation that large carbon grains must be amorphous follows more
from investigation of the structural chemistry in the ISM (Sorrell
1989) than from our radiative-transfer calculations.

The observations become noisier near $5\mum$, which presents a problem
in where to define the continuum level. The calculated continuum at
these short wavelengths is sensitive to the chosen inner radius, which
is constrained by the observed continuum to $(3\hbox{--}5)\times
10^{16}\cm$. The PAH features at 7.7 and $8.6\mum$ are not well
fitted.  For the latter, the SK model consistently predicts values too
low by about a factor $\leq 3$ and it is likely that the cross section
used for this feature is too low.  We use the ones published by
L\'eger et al. 1989a (see also d'Hendecourt et al.  1989, de Muizon et
al. 1990).  The plateau of weaker features between 3.2--3.6 $\mu m$
is at present not treated in the SK dust model.
The feature at $7.0\mu$m, which is not fitted, is due to
an [ArII] line.

For $\lambda \geq 400 \mu m$ contribution from free--free emission
becomes important. The predicted free--free flux, extrapolated from
the radio spectrum, is indicated by the dashed line; the solid line
includes this contribution.

\subsection{ The near infrared }

The J,H,K fluxes contain a large contribution from free--free and
bound--free emission arising from the ionized gas, which is not
included in the model.  We can calculate this contribution from the
known radio flux density, using the tables for a pure hydrogen nebula
in Ferland (1980). From this we predict fluxes of 0.13, 0.11 and
$0.14\,$Jy at J,H,K. Added to this should be the contribution from the
central star: this can be the dominant source for the coolest
planetary-nebulae central stars. For the amorphous-grain model we
predict for the star a flux of 0.40, 0.24 and $0.13\,$Jy at
J,H,K. (The higher luminosity required in the graphite-grain model
leads to somewhat higher values in that model.) The dotted line
indicates the sum of these contributions.

In the amorphous-grain model, we find good agreement at J and H but
some emission deficit at K with respect to the data from Russel et al.
(1977) and a large deficit with respect to the K-flux from Hora et
al. (1993). In the graphite-grain model the K-data from Russel is
fitted better, but now the J and H flux is too large indicating that
this model requires a too-high stellar luminosity.

The reason for the difference between the Russel et al. data and the
Hora et al. data is not clear to us. We note that the flux measured
for the central star by Hora et al. is reproduced reasonably well by
the model ($0.09\,$Jy measured versus $0.13\,$Jy predicted). Willner
et al. (1972) have also claimed an unexplained excess emission at K.

\subsection{ Radial profiles and PAH distribution }

Figure 4 shows radial cross scans through the nebula at several
wavelengths at infinite spatial resolution, as predicted from the final
model parameters for the amorphous-grain model.  In our model, we find
that the radial profile is almost constant with wavelength in the
mid-infrared. Only towards the far-infrared and the submillimetre does
the nebula become much more extended. The absence of change with
wavelength around 10$\mu$m is in part due to the fact that a
significant part of the continuum arises from PAHs, which are also
responsible for the broad band features. Figure 5 shows how the radial
emission profiles of the different dust components behave separately,
at 11.3$\mu$m and 3.3$\mu$m. The PAHs have a larger
extent than either the small graphites or the large grains.
At 3.3$\mu$m the
large grains are not plotted because their contribution becomes
insignificant.

The profiles of Figure 4 can be compared to published infrared maps
(Hora et al. 1990, 1993).  The published profiles have different
extensions in NS and EW direction: at 10$\mu$m, the-IR profiles in
Hora et al. (1993) peak approximately $1.5\arcsec$ away from the
central position in the NS directions, while in EW direction the peak
occurs at about $1.9\arcsec$. This immediately shows that a
spherically symmetric model has limitations for this nebula.
The bottom
panel of Figure  4 shows the profile predicted from our model, smoothed
to $1.2\arcsec$ resolution (where the smoothing was done on a
two-dimensional image derived from the profile). The peak now falls at
about $1.0\arcsec$ from the center, with an uncertainty of about 25\%.
The model predicts therefore a more compact model than actually
observed.  The graphite model predicts a somewhat larger inner radius,
which in fact would better agree with the observations.  However, this
is due to the fact that the graphite model requires a higher stellar
luminosity, which predicts too high a magnitude for the central
star.

Hora et al.  (1993) also find that for BD~+30$^\circ$3639 the PAH
features are more extended than the 10$\mu$m continuum, peaking about
$0.5\arcsec$ further out. This is not seen in our model, and could
only be explained if the dust composition changes with radius. In
fact, Hora et al. suggest that the difference is due to PAH
destruction in the ionized region.  The effect is not always seen in
young planetary nebulae: Meixner et al. (1993) find that the planetary
nebula IRAS21282+5050, which appears to be in in a similar
evolutionary phase as BD~+30$^\circ$3639, shows the same structure at
several mid-infrared wavelengths. In all these cases is the infrared
extent  larger than that of the ionized gas.

By comparing our model with observed radial profiles there are two
disagreements on a sub--arcsecond scale: i) A too-small inner radius,
which may just be a consequence of the simple density distribution
chosen, or alternatively it could be improved by assuming a larger
distance. ii) A different extent of the observed 11.3$\mu$m compared
to the 10$\mu$m emission. This points to a potentially more damaging
problem and may be caused by our neglect of photo-ionization. First,
photo-destruction of PAHs by Ly$\alpha$ is not treated, and second,
the Ly$\alpha$ gives an additional heating source for the grains in
the ionized region. This additional term could slow the decrease in
the radial emission profile for all dust components, but especially
for the smallest grains.  However, the difference in distribution
between 11.3 and 10$\mu$m could only be explained in our model if in
the inner regions the PAHs are partly destroyed. From our preliminary
inclusion of photo-ionization we found that PAH destruction could not
be excluded.

At longer wavelengths the predicted flux distribution becomes more
extended. A consequence of the radial extent of the emission is that
the observed flux depends on the size of the aperture at the
long-wavelength part of the spectrum.  We have calculated the observed
$450\mu$m emission through the $9\arcsec$ and $18\arcsec$ JCMT
apertures, and find 0.15 and $0.28\,$Jy respectively, compared to the
observed values of 0.5 and $0.75\,$Jy (HRC, after subtraction of the
contribution from free--free emission; especially the last value has a
large uncertainty). This shows that, although we used the
submillimetre flux to constrain the outer radius of the density
distribution, in fact the fit here falls short by a factor of three.
The flux depends on both the efficiencies and on the temperature of
the grains.  The temperatures can be calculated in an easy manner; it
is very unlikely that they can be wrong by a large factor. The
efficiencies are more uncertain, and it is possible that the measured
grain properties of Preibisch et al. still underestimate the
submillimetre flux.  Alternatively, the submillimetre flux could have
a contribution from inhomogeneous grain conglomerates, the so-called
``f{}luffy grains'' (Wright 1987, Kr\"ugel \& Siebenmorgen
1994). These grains have higher emissivity by large factors in the
submillimetre regime.

\subsection{ PAH size distribution and hydrogenation}

In the model, the PAH emission is dominated by the largest PAH
component, the PAH clusters. The contribution from the small PAHs is
almost negligible.  This agrees with Hora et al. (1993), who conclude
from a comparison of their 11.2$\mu$m image with a 3.28$\mu$m image by
Roche (1989) that the same component appears to be responsible for
both features.  In order to explain the line ratios, a very low
hydrogenation is needed for the clusters.

In AGB and post-AGB objects, the 11.3$\mu$m is often due to SiC
emission.  This raises the question whether in BD~+30$^\circ$3639 it
also contributes to the observed feature.  Observationally, it is
expected that SiC emission will peak close to the star where PAHs are
probably destroyed. Therefore the data from Hora et al. would favour
PAHs as the origin of the 11.3$\mu$m emission in BD~+30$^\circ$3639,
although is it possible that sub-arcsecond imaging would reveal a SiC
component. It seems likely that
SiC is only a minor contributor to the 11.3$\mu$m (but this could well
be different in other planetary nebulae). We note that identifying the
11.3$\mu$m feature with SiC (e.g.  HRC) would result in an even
lower hydrogenation of the PAHs but also allow for a larger
contribution of small PAHs. In such models the low hydrogenation also
applies to the small PAHs

\subsection{ Internal extinction }

In spite of the low extinction towards
BD~+30$^{\circ}$3639, we predict a significant extinction towards the
central star, caused by the dust inside the nebula. We find $A_{\rm
V}=0.43$ for the amorphous-grain model, and approximately the same for
the graphite model. The extinction towards planetary nebulae is
normally derived from the ratio of the hydrogen lines. However, the
assumption that the extinction towards the star and the nebula are
equal may not be valid for high-density nebulae such as
BD~+30$^{\circ}$3639.  This will affect the commonly used Zanstra
method to determine central-star temperatures, which uses the ratio of
the visual stellar flux over the H$\beta$ flux from the nebula. In the
case of extinction internal to the nebula, the H$\beta$ flux will be
less affected than the stellar flux.  For BD~+30$^\circ$3639,
the effect
could lower the stellar temperature from the Zanstra value of
$30\,000\,$K (used in this paper) to $28\,000\,$K.

The extinction of BD+30$^\circ$3639 has also been measured from
the depth of
the 2200\AA\ absorption feature towards the central star. This
feature is probably caused by small graphites in the interstellar
medium. However, the dust properties in circumstellar mediums are
different and the internal extinction in the nebulae would not
necessarily cause such a feature. Significant internal extinction can
therefore not be ruled out on the basis of presently available
observations. K\"oppen (1977) suggested that internal extinction
should be important for a small number of planetary nebulae, and
derived that especially for  BD~+30$^\circ$3639 the observed
extinction should be almost exclusively caused by the nebula.

\subsection{ Total dust mass and dust-to-gas ratio}

Adding up all the dust components listed in Table 1, we arrive at a
dust-to-gas mass ratio of $3.5\times10^{-3}$ for amorphous grains (and
somewhat larger if graphite grains are assumed). This is smaller than
the ratio found in the ISM ($7\times10^{-3}$ to $10^{-2}$). It should
be noted that our value only applies to the ionised region, since the
gas density in the neutral region is not known. Smaller ratios are
commonly found in PN (e.g., HRC, Barlow 1993) and gives problems if we
assume that the dust in the ISM comes from PN. It is possible that the
ratio in the ionised region is lowered by two competing processes:
dust destruction by UV photons, and dust removal to the outer neutral
region through radiation pressure. In both cases it is expected that
the dust-to-gas ratio will be higher in the neutral region.

The total predicted gas mass is $2.5\msolar$. This value is rather
uncertain: it assumes that the dust-to-gas ratio is constant
throughout the nebula, and it depends on the outer radius which is not
well constrained by our model. However, the large value clearly
favours a high-mass progenitor which was able to shed at least a solar
mass from its atmosphere before leaving the AGB.

\subsection{ The evolutionary status of BD~+30$^\circ$3639}

The model predicts a stellar luminosity of $1.3\times 10^4\lsolar$,
which is consistent with the lower limit found from the radio
emission.  As before, we stress that we used a low value for the
distance, and that the actual luminosity may be even higher.  The
luminosity is high for a PN. The luminosity--core mass relation
(e.g. Boothroyd and Sackman 1988, Pottasch 1992) implies a
corresponding core mass of 0.7--0.75$\msolar$. Most PN have core
masses around $0.6M_{\odot}$, corresponding to luminosities of
approximately $6\times10^3\lsolar$.  Of the well-known galactic PN,
only NGC~7027 and NGC~6369 are known to have such high luminosities
(Gathier and Pottasch 1989), where it should be noted that NGC~7027
appears to have entered the cooling track in the HR diagram, and
therefore will have had a higher luminosity in the past. The high core
mass of BD~+30$^{\circ}$3639 implies a massive progenitor of several
solar masses.

The total nebular mass in our fit of about $2.5\msolar$, although
poorly determined, also implies a massive progenitor of several solar
masses, which has shed a large amount of mass. The density of the
$r^{-2}$ component, together with the known expansion velocity of
$22\kms$ (Acker et al. 1992), yields a mass-loss rate at the AGB of
$2\times10^{-4}\msolar\yr^{-1}$. This is an extremely high rate, with
most observed mass-loss rates on the AGB in the range $2 \times
10^{-5} \hbox{--} 10^{-7} \msolar\yr^{-1}$. It seems likely that the
most massive objects experience the highest mass-loss rates, which
would make the calculated value for BD+30$^{\circ}$3639 acceptable.

 From the inner radius of the nebula, and using the known expansion
velocity, we derive an age of the nebula of around $600\,$yr, measured
from the end of the mass-loss phase.  Sch\"onberner (1989) finds that
for stars with high core mass, the transition time between AGB and
early PN is 500--1000$\,$yr. Therefore, the model results are
self-consistent if BD~+30$^{\circ}$3639 is far more massive than most
galactic PN. Both the evolutionary time scale and the low temperature
of the central star put it in an early phase of a PN evolution.

\section {The infrared evolution of BD+30$^{\circ}$3639}

The density distribution used in the final fit allows one to
calculate the
preceding infrared evolution of BD~+30$^{\circ}$3639. This is
 particularly
interesting because stellar evolution at the phase immediately before
appearance of the visible PN is not well understood. The main
observational reasons are first the high circumstellar extinction,
rendering many objects invisible at optical wavelengths, and second
the fact that for those stars which are optically visible, distance
determinations have not yet been possible. Surveys have been carried
out using the IRAS database, since most of the stellar energy is
converted to circumstellar infrared emission (Volk and Kwok 1989).
The surveys have been partly successful: a number of objects have been
found, but their precise evolutionary phase could not easily be
determined
(e.g. Trams et al. 1991).

In calculating the progenitor evolution we have made a number of
simplifying assumptions. First, the entire nebula was represented only
by the $1/r^{2}$ dust density component. We extended the wind-like
component of the fit in Section 3 inward, keeping the total nebular
mass constant.  This is equivalent to assuming that the circumstellar
envelope of BD~+30$^{\circ}$3639 was formed during a single phase of
constant mass loss. Second, we calculated the age of
BD~+30$^{\circ}$3639 from the adjusted inner radius and the observed
expansion velocity of $22\kms$.  Third, we assumed that the mass loss
terminated at a stellar temperature of $5000\,$K, and that the stellar
temperature has increased linearly during the post-AGB evolution.  The
last assumption is in principle only valid for low and intermediate
core masses. For core masses larger than $0.6\msolar$, the temperature
increase becomes faster after $t=500\,$yr, corresponding to $T_*=
10^4\k$ (Sch\"onberner, 1989).  Therefore, we will somewhat
overestimate the temperature increase in the first $500\,$yr.

The first step of the evolutionary sequence is calculated when the
inner radius reaches $10^{15}\cm$, approximately ten years after the
mass loss ends. In this way we avoid the problem at what distance from
the star the dust particles form, which is a major uncertainty in
modeling on-going mass loss, but not the subject of the present paper.

Figure 6 shows  six evolutionary steps. The corresponding parameter
values are listed in Table 2. The first step shows a steep cut-off
around $2.5\mu$m, caused by the high optical depth in the
circumstellar envelope. At this phase the visual extinction will drop
quickly, and the star will become visible at
near-infrared wavelengths within
$25\,$yr. Optical visibility follows after $\sim 100\,$yr.
The peak flux
of the dust emission is seen to shift to longer wavelengths in
subsequent time steps, with the far-infrared continuum slowly rising
during the evolution. The latter effect is caused by the fact
that the dust
temperature in the outer regions of the cloud increases as the optical
depth towards the star  decreases, giving a stronger radiation
field in the outer layers.

The development of the PAH features is shown in Figure 7a.  They
develop quickly when the star is still relatively cool, and reach full
strength when the star is about $15000\k$.  At this time the star is
still too cool to significantly ionize the nebula: the PAH features
develop before a planetary nebula becomes visible.  Interestingly, the
flux at $3.3\mu$m stays almost constant: the drop in continuum flux is
compensated by the increasing band emission. At later stages (not
shown) the line intensity will go down as the dust moves away from the
star.

In Figure 7b cross scans through the nebula at several wavelengths are
shown.  The submillimetre emission is more extended, especially at
later phases. The distribution implies that aperture effects at these
measurements are much less severe for AGB or very young post-AGB
objects, a prediction that can be easily tested.

 From sequence 3 onwards ($t=360\,$yr), the J,H,K colours are dominated
by the central star, and therefore resemble progressively hotter black
bodies. In contrast, in the L-band dust emission is present at most
stages of evolution, and the object will show an L-excess in
the H$-$K, K$-$L diagram as compared to the black-body line. Only in
the first $\sim100\,$yr are the J,H,K colours influenced by the dust:
in this phase the optical depth at 1--2$\mu$m is still high. This
implies that the J$-$H, H$-$K diagram shows an excess over the
black-body line only for objects which have very recently experienced
mass loss or are still in the mass-loss phase. For more evolved
objects the diagram will show stellar colours only.  We conclude that
the near-infrared colours are not a very sensitive tracer when the
stellar temperatures approach the planetary-nebula phase.

Table 2 lists the IRAS flux densities during the evolution, obtained
by convolving the calculated spectra with the IRAS transmission
functions. The IRAS colours approximate a black body with an
equivalent temperature of $350\k$ in the first model, and evolve
towards the planetary nebula regime.  Figure 8 presents the evolution
of the IRAS colours in the post-mass-loss phase.  Here the colours
evolve significantly during the first 500yr, after which the evolution
slows down.  For comparison the figure also shows the colours of all
Galactic planetary nebulae with good, three-band IRAS detections.  At
the stage that the colour evolution slows down, the colours are
already close to those exhibited by planetary nebulae.  Thus, in order
to find the further evolved ``transition'' objects, surveys should
include objects with infrared IRAS colours close to those of PN.  Most
surveys, however, have concentrated on objects with colours more
similar to AGB stars (e.g., van der Veen, 1988).
Such surveys will have been
biased in favor of young post-AGB objects, or to objects with slower
evolving dust shells, such as found in binary systems.  In order to
find all objects in transition towards planetary nebula, a large range
of IRAS colours is needed.

\section {Summary}

We have used a radiative transfer code for transiently heated dust
particles. Our model include PAH molecules to fit the full IR spectrum
of the planetary nebula BD~+30$^{\circ}$3639. A good fit to the
spectral energy distribution is obtained when we use amorphous carbon
grains for the large particles.  The model does not reproduce the
observed difference between the radial profile at 10$\mu$m and
11.3$\mu$m. This is probably due to the fact that we do not treat
photo-ionization and therefore do not include a Ly$\alpha$ heating
term. The model implies that the observed difference can only be
explained if PAHs are partly destroyed in the inner (ionized) region;
this PAH destruction could also be due to Ly$\alpha$ photons.  In the
submillimetre, the predicted flux is lower than observed in the JCMT
aperture.  This could possibly be explained by a component of of
large, inhomogeneous (``f{}luffy'') dust particles, which have an
enhanced submillimetre emissivity compared to their pure spherical
counter parts.

The result are consistent with a classification of
BD~+30$^{\circ}$3639 as a high-mass planetary nebula, with a core mass
of $\sim 0.7\msolar$, and a mass-loss rate on the AGB of
$2\times10^{-4} M_{\odot} yr^{-1}$. The PAH band ratios imply that the
PAH are mainly in the form of clusters, which are strongly
de-hydrogenated.  Compared to the PAH abundance the contribution of
SiC grains to the 11.3$\mu$m feature is probably small. We find that
the dust-to-gas mass ratio in the ionised region is somewhat smaller
than in the ISM, in agreement with earlier studies (e.g. Hoare et
al. 1992).

We have calculated the infrared spectrum of BD~+30$^{\circ}$3639 at
several stages of its post-AGB evolution. We find that the IRAS
colours of evolved high-mass post-AGB objects($t>500$yr) should
resemble those of planetary nebulae.  Many surveys of post-AGB have
concentrated on objects with IRAS colours intermediate between AGB
stars and planetary nebulae, and may thus have been missed older
post-AGB objects.  Of the near-infrared bands, J,H and K mainly
measure the central star; only the L-band contains significant dust
emission at nearly all evolutionary stages. \\
\\
\\
\\
{\bf Acknowledgements}  We thank A. Tielens for communicating new
data on BD+30$\deg$ prior to publication.
The research has made use of the
SIMBAD database of the Centre d' Astrophysique. We also thank an
anonymous referee for comments.

\pagebreak
\noindent {\bf References}
\begin{list}
{}{\itemsep 0pt \parsep 0pt \leftmargin 3em \itemindent -3em}

\item Acker A., Ochsenbein F., Stenholm B., Tylenda R., Marcout
J., Schon C., 1992, Stras\-bourg--ESO catalogue of galactic planetary
nebulae. European Southern Observatory, Garching

\item Allamandola L.J., Tielens A.G.G.M., Barker J.R., 1985,
ApJ (Letters),
290, L25

\item Allamandola L.J., Tielens A.G.G.M., Barker,J.R., 1989a, ApJS,
71, 733

\item Allamandola L,J., Bregman J.D., Sandford S.A., Tielens A.G.G.M.,
Witteborn F.C., Wooden D.H., 1989b,   ApJ (Letters), 345, L59

\item Barlow M.J., 1993, Wein\-berger R., Acker A., Eds., IAU
Symposium 155,
Planetary Nebulae. Kluwer, Dordrecht,  p.163

\item Basart J.P., Daub C.T., 1987, ApJ, 317, 412

\item Bentley A.F., Hackwell J.A., Grasdalen G.L., Gehrz R.D.,  1984,
ApJ, 278, 665

\item Blanco A., Bussoletti E., Colangeli L., Fonti S., Stephen
J.R., 1991, ApJ (Letters), 382, L97

\item Boothroyd A.I., Sackman I.J., 1988, ApJ, 328, 641

\item Bussoletti E., Colangeli L., Orofino V., 1987, ApJ (Letters),
321, L87

\item Cardelli J.A., Clayton G.C., Mathis J.S., 1989, ApJ, 345, 245


\item D\'esert F.X., Boulanger F., Puget J.L., 1990, A\&A, 237, 215

\item Draine B.T., Lee H.M., 1984, ApJ, 285, 89

\item Draine B.T., Lee H.M., 1987, ApJ, 318, 485

\item Draine B.T., 1988, ApJ, 333, 848

\item Draine B.T., 1989, in: Allamandola L.J., Tielens A.G.G.M., Eds.,
IAU Symp. 135, Interstellar Dust. Kluwer, Dordrecht, p.313


\item Ferland, G.J., 1980, PASP, 92, 596

\item Fitzpatrick E.L., Massa, D., 1988, ApJ, 328, 734

\item Gathier R., Pottasch, S.R., Goss, W.M. 1986, A\&{}A, 157, 191

\item Gathier R., Pottasch, S.R. 1989, A\&{}A, 209, 369

\item Goebel, J.H., Moseley, S.H. 1985, ApJ (Letters), 290, L35


\item Hajian A.R., Terzian Y., Bignell C., 1993, NRAO preprint

\item d'Hendecourt L.B., L\'eger, A., Boissel, P., and Desert, F.,
1989, in: Allamandola L.J., Tielens A.G.G.M., Eds.,
IAU Symp. 135, Interstellar Dust. Kluwer, Dordrecht, p.207

\item Hoare M.G., Roche P.F., Clegg R.E.S., 1992, MNRAS, 258, 257 (HRC)

\item Hora J.L., Deutsch L.K., Hoffmann W.F., Fazio G.G., 1990,
ApJ, 353, 549

\item Hora J.L., Deutsch L.K., Hoffmann W.F., Fazio G.G.,
Shivanandan K., 1993, ApJ, 413, 304

\item Iben I., Jr., Renzini A., 1983, ARA\&A, 21, 271

\item IRAS Low Resolution Catalogue, 1986, A\&{}AS, 65, 607

\item Joblin C., L\'eger A., Martin P., 1992, A\&A (Letters), 393, L79

\item K\"oppen J., 1977,  A\&A, 56, 189

\item  Kr\"ugel E., Siebenmorgen R., 1994, A\&A in press

\item L\'eger A., Puget J.L., 1984, A\&A (Letters), 137, L5

\item L\'eger A., d'Hendecourt L., Defourneau D., 1989a, A\&A, 216,
148

\item L\'eger A., Verstaete L., d'Hendecourt L., et al., 1989b, in:
Allamandola L.J., Tielens A.G.G.M., Eds., IAU Symp. 135,
Interstellar Dust.
Kluwer, Dordrecht, p.173


\item Mathis J.S., 1990, ARA\&A, 28, 37

\item Masson C.R. 1989, ApJ, 346, 243

\item Meixner M., Skinner C.J., Temi P., Rank,D., Bregman J.,
Ball J.R., Keto E., Arens J.F., Jernigan J.G., 1993, ApJ, 411, 266

\item Moseley H., Silverberg, R.F. 1985, in:  Thronsson, H.A.,
Erickson, E.F., Eds. Airborne Astronomy Symposium. NASA
Conf. Publ. 2353, p.233

\item de Muizon M.J., d'Hendecourt L.B., Geballe T.R., 1990, A\&A,
227, 526


\item Pe\~na M., Torres-Peimbert S., 1987, Rev. Mex. Astron. Astrof.,
14, 534

\item Pottasch S.R., 1992, Astronomy \& Astrophysics Review, 4, 215

\item Pottasch S.R., 1984, Planetary Nebulae. Reidel, Dordrecht

\item Preibisch Th., Ossenkopf V., Yorke H.W.,  Henning Th.,
1993, A\&A, 279, 577

\item Puget J.L., L\'eger A., 1989, ARA\&A, 27, 161

\item Roche P.F. 1989, in: Torres-Peimbert S., Ed., IAU Symposium
131, Planetary Nebulae. Kluwer, Dordrecht, p.117

\item Rouleau F., Martin P.G., 1991, ApJ, 377, 526

\item Russel R.W., Soifer B.T., Merrill K.M., 1977, ApJ, 213, 66

\item Sch\"onberner D. 1989, in: Mennessier M.O., Omont A., Eds.,
 From Miras to planetary nebulae: which path for stellar evolution?
Editions Fronti\`eres, Gif-s\^ur-Yvette, p.533

\item Schutte W.A., Tielens A.G.G.M., 1993, ApJ, in press


\item Siebenmorgen R., Kr\"ugel E.,1992, A\&A, 259, 614 (SK)

\item Siebenmorgen R., Kr\"ugel E., Mathis J.S., 1992, A\&A, 266,
501

\item Siebenmorgen R., 1993, ApJ ApJ 408, 218

\item Sorrell W.H., 1989, MNRAS, 214, 89

\item Taylor A.R., Pottasch S.R., Zhang C.Y., 1987, A\&{}A, 171, 178


\item Tokunaga A.T., Sellgren K., Smith R.G., et al., 1991,
ApJ, 380, 452

\item Trams N.R., Waters L.B.F.M., Lamers H.J.G.L.M., Waelkens C.,
Geballe T.R., Th\'e P.S., 1991, A\&{}AS, 87, 361

\item Volk K.M., Kwok S., 1989, ApJ, 342, 345

\item van der Veen W.E.C.J. 1988, PhD thesis, University of Leiden

\item Weidemann V.,  Sch\"onberner D., 1989, in: Mennessier M.O.,
Omont A., Eds., From Miras to Planetary Nebulae: Which path for
stellar evolution?  Editions Fronti\`eres, Gif-s\^ur-Yvette, p. 3.

\item Willner S., Becklin E.E., Visnavathan N., 1972, ApJ, 175, 699


\item Wright, E., 1987, ApJ 320, 818

\item Zijlstra A.A., Pottasch S.R. 1989, A\&{}A, 216, 245

\end{list}

\pagebreak

\noindent {\bf  Figure Captions}
\begin{list}
{}{\itemsep 0pt \parsep 0pt \leftmargin 3em \itemindent -3em}

\item Figure 1: Contribution of the different dust components to
the calculated spectrum.  The dashed line represent the model with
large amorphous carbon grains only. The contribution
from large grains and
small graphites is given by the dotted line and, the model including
PAHs is shown by the full line.

\item Figure 2: a) The spectral energy distribution of
BD~+30$^{\circ}$3639. Observations are shown by squares, the model
with amorphous carbon grains for the large-grain component is depicted
by a solid line. the triangles present data from Russel et
al. (1977). In the submillimetre the contribution of free--free
emission (separately indicated by the dashed line) has been added.
The dashed line in the near infrared includes the contribution from
bound--free and free--free emission to the J,H,K,L. The lowest K-point
is a measurement of the star only.  b) High-resolution spectrum of
BD~+30$^{\circ}$3639 between 3 and 14$\mu$m.  Observations shown as
small circles are from Schutte et al. (1993) and the full line is the
model fit.

\item Figure 3: As Figure 2, but now the fit is derived with
Draine \& Lee
optical constants for the large-grain component.

\item Figure 4: a) The normalized radial surface-brightness
distribution
across the nebula at several wavelengths, for the amorphous-grain
model. The top axis shows the radius in arcseconds. b) The 10 $\mu m$
model profile smoothed to 1.2" resolution, as in a). Smoothing
was done
with a Gaussian on a two--dimensional image derived from the model
profile.

\item Figure 5:
The distribution of the individual dust components for the radial
profiles at 3.3$\mu$m (upper panel) and 11.3$\mu$m (lower panel).
The large grains are not shown
for the 3.3$\mu$m because they do not give a significant contribution.

\item Figure 6: The infrared time evolution
between 0.1 and 1000$\mu$m of
BD~+30$^{\circ}$3639. The first evolutionary sequence is calculated 13
years after the mass loss ends. The contribution from radio emission
is not included.

\item Figure 7: a) The evolution of the PAH bands between 13 and 900
years after the mass loss ends. b) The radial surface-brightness
distribution across the nebula at the time steps of Figure 6.

\item Figure 8: The calculated evolution of the IRAS colours for the
time steps of Figure 6. The circles show the actual colours of the
Galactic planetary nebulae with good three-band detections where the
position of the IRAS source and the planetary nebula agree within
$15\arcsec$.
\end{list}

\pagebreak

\begin{table}
\small
{\vbox {Table 1: Model parameters of BD~+30$^{\circ}$3639}
{\noindent \halign{ # \hfil &\quad # \hfil & # \hfil &
\quad# \hfil & \quad # \hfil &  # \hfil \cr
 \noalign{\medskip\hrule\medskip}

Model          &       I                 &                 II
         &    &      I    &    II   \cr
               & (amorphous)             &             (graphite)
         &   & (amorphous)             &             (graphite)
\cr
 \noalign{\medskip\hrule\medskip}
     \multispan3  {\it Dust parameters} \hfill
  & \multispan3  {\it Nebular parameters}\hfill
\cr
$M_{\rm d}/M_{\rm g}$ & $3.5\times 10^{-3}$  &    $9.2\times10^{-3}$ &
   D (kpc) &   2                 &         2
\cr
{\it Large  carbon grains} & & &
 $T_*$\ (K)     &   $3\times10^4$         &          $3\times10^4$
\cr
Lower size (\AA)            &300          &   300    &
  $L_*$\ ($\lsolar$)     &   $1.3\times10^4$  & $ 3\times10^4$
\cr
Upper size (\AA)      & 1200              &  1200
 & $A_1$ & $2\times 10^4$      &         $2\times10^4$
\cr
Abundance & $8.4 \times 10^{-5}$ & $2.4 \times 10^{-4}$
& $A_2$ & $2.3\times10^{38}$ &  $6\times10^{37}$
\cr
{\it Small graphites}  & &
& $R_{\rm i}$ (cm) &   $4\times 10^{16}$    &         $6\times10^{16}$
\cr
Lower size (\AA) &  10                  &    10
& $R_{\rm o}$ (cm)  &    $2.5\times 10^{18}$  & $2.5\times10^{18}$
\cr
Upper size (\AA) &  80                  &    80
& $R_2$ (cm)      &   $8\times 10^{16}$    &         $8\times10^{16}$
\cr
Abundance & $2 \times 10^{-5}$ & $6 \times 10^{-5}$
\cr
{\it Small PAH} \cr
Number of C-atoms  &     25                    &   25
\cr
Abundance          &   $7\times10^{-8}$      &   $10^{-8}$
\cr
H/C atom ratio     &     0.2                   &   0.1
\cr
{\it PAH clusters} \cr
Number of C-atoms   &    250                   &   250 \cr
Abundance           &    $5\times10^{-6}$    &   $10^{-6}$ \cr
H/C atom ratio      &    0.04                  &   0.2 \cr
visual extinction $A_{\rm V}$ & 0.43           &   0.45 \cr
 \noalign{\medskip\hrule\medskip}
}}

$M_{\rm d}/M_{\rm g}$ is derived form the calculated dust density by
assuming an electron density of $2 \times 10^4\,\rm cm^{-3}$. The
abundance of the various dust components is defined as the number of
C-atoms locked up in dust relative to the number of H-atoms, and is
calculated using the same electron density}
\end{table}

\begin{table}
{\vbox {Table 2: The post-AGB evolution of the progenitor of
BD~+3030$^{\circ}$3639}
{\noindent \halign{ # \hfil \quad & # \hfil \quad &# \hfil \quad &
# \hfil \quad & # \hfil \quad &# \hfil &# \hfil
& # \hfil &#  \cr
\noalign{\medskip\hrule\medskip}
 $T_*$ & $R_{\rm i}$ & $t$ & $A_{\rm
V}$ & $F_{12}$ & $F_{25}$ & $F_{60}$ & $F_{100}$ & $F_{450}$ \cr (K) &
($10^{16}\cm$) & (yr) & & \multispan5 (Jy) \cr
\noalign{\medskip\hrule\medskip}
 5000   &    0.1     &       13  &   29.6    &     225 & 165 & 52  &
18 &   0.24 \cr
10000   &    1.3     &      180  &    2.3    &
217 & 262 & 104 & 37 &  0.38 \cr
15000   &    2.6     &      360  &    1.1    &
161 & 284 & 141 & 51 &  0.47 \cr
20000   &    3.9     &      540  &    0.74   &
124 & 270 & 157 & 57 &  0.51 \cr
25000   &    5.2     &      720  &    0.55   &
100 & 249 & 163 & 61 &  0.53 \cr
30000   &    6.5     &      900  &    0.43   &
78  & 225 & 164 & 62 &  0.55 \cr
\noalign{\medskip\hrule\medskip}
}}
\small
The flux densities at 12--100$\mu$m are derived by convolving with the
IRAS bandpasses. the 450$\mu$m is a narrow-band flux, but not
corrected for aperture effects such as would be
measured at JCMT. In the last few models the actual observed 450$\mu$m
flux would be higher due to free--free emission from the ionized gas.
The time is calculated assuming a constant expansion velocity of
$22\kms$.  }
\end{table}

\end{document}